
\font\twelvebf=cmbx12
\font\ninerm=cmr9
\nopagenumbers
\magnification =\magstep 1
\overfullrule=0pt
\baselineskip=18pt
\line{\hfil CCNY-HEP 3/95}
\line{\hfil March 1995}
\vskip .8in
\centerline{\twelvebf Curiosities on Free Fock Spaces}
\vskip .5in
\centerline{\ninerm D. MINIC }
\vskip .1in
\centerline{ Physics Department}
\centerline{City College of the City University of New York}
\centerline{New York, New York 10031.}
\centerline {E-mail: minic@scisun.sci.ccny.cuny.edu}
\centerline{ }
\vskip 1in
\baselineskip=16pt
\centerline{\bf Abstract}
\vskip .1in

We consider some curious aspects of single-species free Fock spaces,
such as novel bosonization and fermionization formulae and relations to various
physical properties of bosonic particles. We comment on generalizations
of these properties to physically more interesting many-species free Fock
spaces.

\vfill\eject

\footline={\hss\tenrm\folio\hss}

\def \12 {{\textstyle {1\over 2}}}

\magnification =\magstep 1
\overfullrule=0pt
\baselineskip=22pt
\pageno=2
\vskip .2in
1. Introduction and background: The old idea of master
fields in large N gauge
theories [1] has been recently revived by Douglas [2],[3] and Gopakumar and
Gross [4]. It appears that the master field concept is naturally
described in the framework of non-commutative probability theory
[5]. One of the crucial ingredients of non-commutative probability theory is
the idea of
free Fock spaces [5], [6], [7]. The free Fock space is  defined
to be a Hilbert space with   orthonormal basis vectors on which
strings
 of creation ($a^{\dagger}_{i}$) and annihilation operators ($a_{i}$) act,
satisfying no relations whatsoever. Hence the adjective free. The basic
operations of $a^{\dagger}_{i}$ and $a_{i}$ are given by
$$\eqalign{
a^{\dagger}_{p}|p_{1},...,p_{n}>~&=|p,p_{1},...,p_{n}> \cr
a_{p}|p_{1},...,p_{n}>~&=\delta_{p,p_{1}}|p_{2},...,p_{n}>, \cr} \eqno(1)
$$
where $a|0>=0$. It follows then that
$a_j a^{\dagger}_k =\delta_{jk}$ (this relation defines
the so called Cuntz
algebra). The same expression also defines the so called
infinite statistics [8], in which any representation of the symmetric
group can exist. Free Fock spaces appear to have many unusual properties.
For example, it was
argued by Greenberg [8]
that if one attempts to use infinite statistics in order
to  construct  a second-quantized relativistic field theory of free fields, the
resulting theory does not possess
the property of locality and there is no analog of spin-statistics theorem,
even
though CPT theorem and cluster decomposition  seem to hold.
Note as well that from  the Cuntz algebra structure  there appears not to exist
the usual  classical limit, with the well defined
classical phase-space structure.

In order to understand the mechanics
of the free Fock space it is important to address some basic questions.
In the previous note [9] we  asked what  the naive analog of the Gaussian
coherent states  is for the case of infinite statistics. This was motivated by
the fact that the knowledge of the
$N=\infty$ master field  is equivalent to the knowledge of the ground state of
the $N=\infty$ theory [10] and the fact that Gaussian coherent states describe
rather well the ground states of large N vector models.
Unfortunately, the naive analog
of Gaussian coherent states constructed in [9] turns out to possess rather
complicated properties. It is natural to ask next if one can work directly with
Lagrangians and path integrals, appropriately constructed for the case of free
Fock spaces, by taking as a working  hypothesis the notion that the
latter  describe reasonably well
the non-perturbative Fock spaces of the large N
matrix models. (Recently Migdal has addressed a similar
question of construction
of second quantized $N=\infty$
effective planar theory [11], albeit within the framework of momentum loop
equations.)

In this note we consider certain peculiar properties of single-species
free Fock spaces.
In particular we construct an analog of the free harmonic oscillator, and
examine the resulting path integral. We resolve an apparent
puzzle presented by
the resulting path integral (namely that it describes a free bosonic particle,
even though the underlying commutation relations are not bosonic), by
pointing out  the existence of novel  bosonization as well as fermionization
formulae and find out how ordinary Gaussian coherent states fit into this
picture. We also examine, as an extra check,
the quantum partition function for this
system, and find that it corresponds to the well known
result for the bosonic
harmonic oscillator. We discuss generalizations of the
above mentioned results to
physically
more relevant many-species free Fock spaces, having applications to
realistic large N matrix models.
\vskip.2in
2. Single-species spaces:
Consider the number operator $N$, defined by, as usual,
$[N,a^{\dagger}]=a^{\dagger}$, for
a slightly more general "deformed" commutation relation
$aa^{\dagger}-qa^{\dagger}a=1$. (Infinite statistics formally corresponds to
the $q\rightarrow 0$ limit.) Insert the following ansatz
$$
N=\sum_{i=1}^{\infty}c_{i}(a^{\dagger})^{i}a^{i}. \eqno(2)
$$
Then it is easy to show that
$$
c_{n}={(1-q)^{n} \over {1-q^{n}}} .\eqno(3)
$$
Here $(a^{\dagger})^{n}|0>=\sqrt{1(1+q)...(1+q+...+q^{n-1})}|n>$.

Let us examine the Hamiltonian $H=N$. We take this to define  an analog of the
harmonic oscillator Hamiltonian.
Given the definition of the Gaussian coherent states
$$
|z>=\exp(-{|z|^{2} \over 2})\sum_{n}{z^{n} \over \sqrt{n!}} |n>,  \eqno(4)
$$
with the usual properties of the resolution of unity
$$
{1 \over \pi} \int d^{2}z |z><z|=1
$$
and exponential overlap
$$
<z'|z>=\exp(--{|z'|^{2} \over 2}-{|z|^{2} \over 2}+\bar{z'}z),
$$
it follows
that $<z'|N|z>=\bar{z}'z<z'|z>$. Note that we cannot write
$<z'|N|z>=N(\bar{z}',z)<z'|z>$, as the Gaussian coherent states are not
eigenstates of $a$.
Let us consider the propagator
$$
<z_{f}|\exp(-iNt)|z_{i}>.
$$
Using, as usual,
the resolution of unity and the exponential overlap of
the Gaussian coherent
states, we obtain
$$
<z_{f}|\exp(-iNt)|z_{i}>=\int
D z D \bar{z}
\exp \left( i\int_{0}^{t} dt \left[{1 \over 2i}({d\bar{z}
\over dt}z - \bar{z}{dz \over dt}) -\bar{z}z \right] \right), \eqno(5)
$$
which corresponds to the well-known holomorphic
path integral representation of
the bosonic harmonic oscillator.
We are now faced with a slight puzzle. Namely, it appears that there exists an
infinite ambiguity in quantization: for any $q$ ,
say, between zero and one, the
 path integral of the analog of the free harmonic oscillator looks the same. If
we insist that
there should be no relations between creation and annihilation operators
of different index (see the comment after eq. (12)),
then $q=0$ case is singled out.
 Thus we concentrate on infinite statistics in what
follows.

The resolution of the puzzle is presented by a peculiar looking
bosonization formula.
Let
$$
b=\sum_{i=1}^{\infty}\alpha_{i}(a^{\dagger})^{i-1}a^{i} . \eqno(6)
$$
Then it can be shown that $[b,b^{\dagger}]=1$ if
$$
\alpha_{n}= \pm \sqrt{n} \mp \sqrt{n-1} . \eqno(7)
$$
Moreover, given these values for $\alpha_{n}$ it follows that
the number operators corresponding to bosonic and infinite statistics
are the same, namely
$N_{b}=N_{i}$ where, $N_{b}=b^{\dagger}b$ and
$N_{i}=\sum_{i=1}^{\infty}(a^{\dagger})^{i}a^{i}$.
Note also that the Gaussian coherent states which are eigenstates
of $b$ ($b|z>=z|z>$) are eigenstates of a very complicated infinite
combination of $a$ and $a^{\dagger}$, defined by (6).

It is true as well, that the quantum partition function for
a system of noninteracting
harmonic oscillators $Z=Tr\exp(-\beta H)$, with the
hamiltonian $H=N$, is given by
the well known partition function of the
ideal Bose-Einstein gas. The resulting
 distribution function is  therefore Planckian.

It is possible to construct fermionic operators in a similar manner.
Let
$$
f=\sum_{i=1}^{\infty}\beta_{i}(a^{\dagger})^{i-1}a^{i}.  \eqno(8)
$$
Then it can be likewise shown that $\{ f,f^{\dagger} \}=1$ provided
$$
\beta_{2n-1}=\pm 1, ~~ \beta_{2n}=-\beta_{2n-1}  \eqno(9)
$$
for $n=1,2,...$.
One can also check that $f^{2}={f^{\dagger}}^{2}=0$; hence these are genuine
fermion variables.
Note that the above formulae differ from the well-known Jordan-Wigner type of
bosonization formulae.

As an aside we quote the most general expressions, that interpolate
between formulae (6) and (8) and (7) and (9). Let
$$
c = \sum_{i=1}^{\infty} \gamma_{i} (a^{\dagger})^{i-1} a^{i}. \eqno(10)
$$
Then $c c^{\dagger} - q c^{\dagger} c = 1$ if
$$
\gamma_{i} = \mp \sqrt{1+q+...+q^{i-2}} \pm \sqrt{1+q+...+q^{i-1}}.  \eqno(11)
$$
Here $i=2,3,...$ and $\gamma_{1}=\pm 1$.

Formulae (6)-(9) do not look that strange if we remember that any
representation of the symmetric group can exist for infinite statistics.
It is not surprising therefore that there exist  realizations
of symmetric and antisymmetric representations in terms of operators
satisfying infinite statistics.
\vskip.2in
3. Many-species spaces:
Can we generalize above formulae to the case of many-species spaces?

First of all we can immediately write down the general expression for the
number operator $N_{i}$ (see [8])
$$
N_{i}=\sum_{j=1}^{\infty}\sum_{k_{1},...,k_{j-1}}
a^{\dagger}_{k_{1}}...a^{\dagger}_{k_{j-1}}
a^{\dagger}_{i}a_{i}a_{k_{j-1}}...a_{k_{1}}.  \eqno(12)
$$
Note that in the case of infinite statistics, this expression follows
without any assumptions about relations between creation and annihilation
operators of different index. This is not the case if
$a_{i}a^{\dagger}_{j} - q a^{\dagger}_{j}a_{i}= \delta_{ij}$ and $q \neq 0$.

What about the bosonization and fermionization formulae?
If we demand
$$
b_{i}=\sum_{j=1}^{\infty}\sum_{k_{1}...k_{j-1}}\alpha_{j}
a^{\dagger}_{k_{1}}...a^{\dagger}_{k_{j-1}}
a_{k_{j-1}}...a_{k_{1}}a_{i} \eqno(13)
$$
then the following relation holds
$$
b_{i}b^{\dagger}_{j}=\delta_{ij} a_{j}b_{j}^{\dagger}b_{i}a_{i}^{\dagger},
\eqno(14)
$$
which indeed reduces to the previously quoted expressions for $i=1$. So the
 bosonization and fermionization formulae, defined through (13), are
true only for $i=1$.

Likewise, one can examine the quantum partition function for this generalized
situation. This can be easily accomplished if we remember that there
exist no relations between $a_{i}$ and $a_{i}^{\dagger}$; therefore the
relevant combinatorial factor is Boltzmannian. The resulting quantum partition
function is then (this result was established in  conversations with
V.P.Nair)
$$
Z=(1 - \sum_{k} \exp(-\beta E_{k}))^{-1}.  \eqno(15)
$$
Here $E_{k}$ denotes the appropriate energy eigenvalue.
In the $k=1$ limit we recover
the previous result. Note that in this general case, the quantum
partition function
is not always well defined. In particular the large volume limit is
singular and the expectation value for the number
operator is not always positive
definite. Note also that the purely classical partition function is of
the Boltzmann type except for the omission of the Gibbs ${1 \over N!}$ factor,
and is therefore well defined (see [8]). In view of this observation and
of Wigner's original work on  random matrices, it seems that
instead of the quantum partition function the appropriate physical quantity
one should  consider is   the classical Gibbsian partition
function, though appropriately "projected" [12].

Thus we clearly see that results valid for the single-species case are
 very special.
\vskip.2in
4. Conclusions:
We have examined some basic properties of single-species free Fock
spaces, in particular the path integral
(and the quantum partition function) of an
analog of the free harmonic oscillator and found that
it corresponds to the usual
path integral (and the quantum partition function)
of a free bosonic particle. This
slightly puzzling result was understood by utilizing
a bit peculiar looking
bosonization formula. (Similar fermionization formula was discussed as well.)
These results do not generalize to many-species spaces indicating
that properties of many-species free Fock spaces
appear not to be straightforward
extensions of the single-species case (as we are usually
accustomed to). It is of utmost importance to understand these, before we
can even attempt to address the physics behind realistic  large N theories.

\vskip.2in
I greatfully acknowledge so many helpful discussions with V.P.Nair that
have shaped this note. This work was supported in part by
the NSF grant PHY 90-20495 and by the  Professional Staff Board of Higher
Education of the City University of New York
under grant no. 6-63351.
\vskip.1in
{\bf References}
\item{1} E.Witten, in {\sl Recent Developments in Gauge Theories}, eds.
G.'tHooft et al., Plenum Press, New York and London (1980).
\item{2.} M.R.Douglas, "Large N Gauge theory - Expansions and Transitions", to
appear in the proceedings of the 1994 ICTP Spring School, hep-th/9409098.
\item{3.} M.R.Douglas, "Stochastic Master Fields", RU-94-81, hep-th/9411025;
M.R.Douglas and M.Li, "Free Variables and the Two Matrix Model", BROWN-HET-976,
RU-94-89, hep-th/9412203.
\item{4.} R.Gopakumar and D.J.Gross, "Mastering the Master Field", PUPT-1520,
hep-th/9411021.
\item{5.} D.V.Voiculescu, K.J.Dykema and A.Nica, {\sl Free Random Variables},
AMS, Providence (1992).
\item{6.} O.Haan, Z.Phys C6 (1980) 345; see also M.B.Halpern and C.Schwarz,
Phys.Rev. D24 (1981) 2146 and A.Jevicki and H.Levine, Ann.Phys. 136 (1981)
113.
\item{7.} P.Cvitanovi\'{c}, Phys.Lett. 99B (1981);
P.Cvitanovi\'{c}, P.G.Lauwers
and P.N.Scharbach, Nucl.Phys. B203 (1982) 385.
\item{8.} O.Greenberg, Phys.Rev.Lett. 64 (1990) 705.
\item{9.} D.Minic, "Remarks on Large N Coherent States", CCNY-HEP 1/95,
hep-th/95002117.
\item{10.} L.G.Yaffe, Rev.Mod.Phys. 54 (1982) 407; F.Brown and L.G.Yaffe,
Nucl.Phys. B271 (1986) 267; T.A.Dickens, U.J.Lindqwister, W.R.Somsky and
L.G.Yaffe, Nucl.Phys. B309 (1988) 1.
\item{11.} A.Migdal, "Second Quantization of the Wilson Loop", PUPT-1509,
hep-th/9411100.
\item{12.} M.L.Mehta: {\sl Random Matrices}, Academic Press,
New-York and London (1967).

\end